\documentclass[11pt]{article}

\usepackage[a4paper,margin=1in]{geometry}
\usepackage{amsmath,amssymb,mathtools}
\usepackage{bm}
\usepackage{hyperref}
\usepackage{cite}

\hypersetup{
  colorlinks=true,
  linkcolor=blue,
  citecolor=blue,
  urlcolor=blue,
  pdftitle={The Constructive Cavity Method for the Ising Perceptron},
  pdfauthor={Simone Franchini, Francesco Concetti, Giampiero Bardella}
}

\newcommand{\E}{\mathbb E}
\newcommand{\R}{\mathbb R}
\newcommand{\dd}{\,\mathrm d}
\newcommand{\Dz}{\mathrm Dz}
\newcommand{\one}{\mathbf 1}
\newcommand{\mcG}{\mathcal G}
\newcommand{\mcA}{\mathcal A}
\newcommand{\mcH}{\mathcal H}
\newcommand{\eps}{\varepsilon}

\title{The Constructive Cavity Method for the Ising Perceptron}

\author{%
\begin{tabular}{c}
Simone Franchini\textsuperscript{1,*}, Francesco Concetti\textsuperscript{2}\\
Giampiero Bardella\textsuperscript{3}\\[0.6em]
\small \textsuperscript{1}~CNR-ISTC, Via Gian Domenico Romagnosi 18, 00196 Rome, Italy.\\
\small \textsuperscript{2}~Faculty of Mathematics and Computer Science, UniDistance Suisse, 3900 Brig, Switzerland.\\
\small \textsuperscript{3}~Department of Physiology and Pharmacology, Sapienza University of Rome,\\
\small Piazzale Aldo Moro 5, 00185 Roma, Italy.\\[0.3em]
\small \textsuperscript{*}~Contact: \href{mailto:simone.franchini@yahoo.it}{simone.franchini@yahoo.it}
\end{tabular}%
}
\date{}

\begin{document}

\maketitle

\begin{abstract}
We apply the Constructive Cavity Method (CCM) to the Ising perceptron.  The process starts from positive soft channels admitting an auxiliary representation
\[
U(x)=\int_{\mathbb R} e^{tx}\rho(\dd t),\qquad \rho\ge0,
\]
with an auxiliary measure having a finite exponential quadratic moment. This rewrites the perceptron as a dense, bipartite Gaussian model with an Ising layer and a continuous auxiliary layer.  The CCM blanket geometry then gives two constructive variance profiles, \(Q_\ell\) on the Ising side and \(\Gamma_\ell\) on the auxiliary side.  The bipartite shell scale is
\[
Q_{\ell+1}\Gamma_{\ell+1}-Q_\ell\Gamma_\ell,
\]
and Gaussian core comparison removes the core contribution \(\Delta Q_\ell\Delta\Gamma_\ell\) under blanket refinement.  The scalar cavity functional has three terms,
\[
\mathcal A_{\mathrm{perc}}(\nu)
=
\mathbb E\left[
\log\langle2\cosh x\rangle_\nu
+\alpha\log\langle U(s)\rangle_\nu
-\log\langle e^y\rangle_\nu
\right],
\]
where \(x\) is the new-spin field, \(s\) is the new-constraint field, and \(y\) is the cavity correction.  CCM identifies the constructive cavity variables with the Aizenman--Sims--Starr random overlap structure (ROSt) variables. Parisi's recursive averaging then gives
\[
\mathcal G_{\mathrm{perc}}^{\mathrm{CCM}}(\zeta,Q,\Gamma)
=
X_0(0)+\alpha Y_0(0)
-\frac12\sum_{\ell=0}^{R}\zeta_\ell
\left(Q_{\ell+1}\Gamma_{\ell+1}-Q_\ell\Gamma_\ell\right).
\]
The replica-symmetric specialization gives Gardner's formula for the generalized Ising perceptron.  Specializing the constraint terminal to the hard channel gives the Krauth--M\'ezard functional for the binary perceptron.
\end{abstract}

\newpage

\section{Introduction}

The Constructive Cavity Method (CCM) is a computation method that identifies the constructive cavity variables with the Aizenman--Sims--Starr random overlap structure (ROSt) variables and evaluates the cavity functional by the Parisi's recursive averaging~\cite{SimplifiedParisiIII}. The procedure applies the blanket construction of \cite{FranchiniRSBWithoutReplicas,FranchiniMartingale,FranchiniSimplifiedParisi} and REM universality~\cite{FranchiniREMUniversality,ConcettiFranchiniLinearREM,ConcettiFranchiniLinearREMLetter}. The corresponding computation for dense Gaussian RBM/DBM architectures is developed in \cite{FranchiniConcettiBardellaDBM}. We use the RBM/DBM terminology in the standard sense of a Boltzmann-machine model \cite{AckleyHintonSejnowskiBM,SmolenskyHarmony,SalakhutdinovHintonDBM}, and in relation with the multi-species and DBM spin-glass literature \cite{BarraGenoveseSollichTantariRBM,BarraContucciMingioneTantariMultispecies,PanchenkoMultiSpecies,AlbericiContucciMingioneDBM,GenoveseMinimaxDBM,AgliariDBM}.

The aim of this paper is to apply the same operational calculation to the Ising perceptron. We consider the partition function
\begin{equation}
Z_N
=
\sum_{\sigma\in\{-1,+1\}^N}
\prod_{\mu=1}^{M_N}
U\left(
\frac1{\sqrt N}
\sum_{i=1}^N \xi_i^\mu\sigma_i
\right),
\qquad
M_N=\alpha N+o(N),
\label{eq:intro-perceptron-model}
\end{equation}
where the patterns \(\xi_i^\mu\) are independent standard Gaussians and \(U\ge0\) is the channel. For the positive-channel calculation we assume that
\begin{equation}
U(x)=\int_{\R}e^{tx}\rho(\dd t),
\qquad
\rho\ge0,
\label{eq:intro-positive-channel}
\end{equation}
under the integrability condition \eqref{eq:auxiliary-integrability}. Then \eqref{eq:intro-perceptron-model} can be written as
\begin{equation}
Z_N
=
\sum_{\sigma\in\{-1,+1\}^N}
\int\prod_{\mu=1}^{M_N}\rho(\dd t_\mu)
\exp\left\{
\frac1{\sqrt N}
\sum_{i=1}^N\sum_{\mu=1}^{M_N}
\xi_i^\mu\sigma_i t_\mu
\right\}.
\label{eq:intro-auxiliary-partition}
\end{equation}
The perceptron is thereby rewritten as a dense, bipartite, Gaussian model with one Ising layer and one continuous auxiliary layer.

This auxiliary representation is the step that places the perceptron into the CCM class. In the spin-only representation a blanket increment enters the nonlinear factor \(U(S_\mu)\), with
\begin{equation}
S_\mu(\sigma)=\frac1{\sqrt N}\sum_i\xi_i^\mu\sigma_i,
\end{equation}
and its logarithmic increment need not be a one-body function of the new spins. In this auxiliary representation the exponent is bilinear. Once the past is fixed, newly added variables in one layer see Gaussian fields generated by old variables in the other layer. This is the same interface mechanism used in the RBM/DBM calculation \cite{FranchiniConcettiBardellaDBM}.

The construction uses two variance profiles. We will write
\begin{equation}
0=Q_0\le Q_1\le\cdots\le Q_{R+1}=1
\end{equation}
for the Ising spin side and
\begin{equation}
0=\Gamma_0\le \Gamma_1\le\cdots\le \Gamma_R<\infty,
\qquad
\Gamma_{R+1}:=\Gamma_R,
\end{equation}
for the auxiliary variables side. The symbol \(\Gamma\) will denote the auxiliary-side constructive variance profile. In the perceptron literature the corresponding conjugate parameter is often denoted by \(\widehat q\); here that notation is used only in the comparison with Gardner--Krauth--M\'ezard formulas. Core terms are denoted explicitly by \(\mcH_\ell^{\mathrm{core}}\).
The bipartite cumulative scale is
\begin{equation}
Q_\ell\Gamma_\ell.
\end{equation}
Thus the shell increment at level \(\ell\) is
\begin{equation}
\Delta(Q\Gamma)_\ell
=
Q_{\ell+1}\Gamma_{\ell+1}-Q_\ell\Gamma_\ell.
\end{equation}
Introducing a notation for the difference
\begin{equation}
\Delta Q_\ell=Q_{\ell+1}-Q_\ell,
\qquad
\Delta\Gamma_\ell=\Gamma_{\ell+1}-\Gamma_\ell,
\end{equation}
one obtains that the shell increment is
\begin{equation}
\Delta(Q\Gamma)_\ell
=
Q_\ell\Delta\Gamma_\ell
+
\Gamma_\ell\Delta Q_\ell
+
\Delta Q_\ell\Delta\Gamma_\ell.
\label{eq:intro-interface-core-split}
\end{equation}
The first two terms are the interface scale. The last term is the core scale. Refining the blankets suppresses the total core scale,
\begin{equation}
\sum_\ell \Delta Q_\ell\Delta\Gamma_\ell\longrightarrow0,
\end{equation}
and the Gaussian comparison \eqref{eq:core-pressure-bound} gives the interface-only pressure in this limit.

The scalar cavity functional associated with the auxiliary perceptron has the form
\begin{equation}
\mcA_{\mathrm{perc}}(\nu)
=
\E\left[
\log\langle2\cosh x\rangle_\nu
+
\alpha\log\langle U(s)\rangle_\nu
-
\log\langle e^y\rangle_\nu
\right].
\label{eq:intro-cavity-functional}
\end{equation}
The field \(x\) is produced by adding one Ising spin, \(s\) by adding one new perceptron constraint, and \(y\) is the cavity correction. Their constructive decompositions are
\begin{equation}
x^{\mathbf a}
=
\sum_{\ell=0}^R
\sqrt{\Gamma_{\ell+1}-\Gamma_\ell}\,z_\ell(\mathbf a_{\le\ell}),
\end{equation}
\begin{equation}
s^{\mathbf a}
=
\sum_{\ell=0}^R
\sqrt{Q_{\ell+1}-Q_\ell}\,\eta_\ell(\mathbf a_{\le\ell}),
\end{equation}
for the constructive cavity variables and
\begin{equation}
y^{\mathbf a}
=
\sum_{\ell=0}^R
\sqrt{Q_{\ell+1}\Gamma_{\ell+1}-Q_\ell\Gamma_\ell}\,g_\ell(\mathbf a_{\le\ell}),
\end{equation}
for the fugacity correction, with the hierarchical labels and Gaussian families defined in Section~\ref{sec:constructive-marks}.
Then, let
\begin{equation}
0=\zeta_0\le \zeta_1\le\cdots\le\zeta_R=1
\end{equation}
be the Parisi averaging parameters. For
\begin{equation}
T_{m,a}f(u)
=
\frac1m\log\E_Z\exp\{m f(u+\sqrt a Z)\},
\qquad
T_{0,a}f(u)=\E_Z f(u+\sqrt a Z),
\end{equation}
the site recursion starts from the initial condition
\begin{equation}
X_{R+1}(u)=\log2\cosh u
\end{equation}
and satisfies the recursive relation
\begin{equation}
X_\ell
=
T_{\zeta_\ell,\Gamma_{\ell+1}-\Gamma_\ell}X_{\ell+1}.
\end{equation}
The constraint recursion starts from
\begin{equation}
Y_{R+1}(u)=\log U(u)
\end{equation}
and also satisfies the same recursive relation
\begin{equation}
Y_\ell
=
T_{\zeta_\ell,Q_{\ell+1}-Q_\ell}Y_{\ell+1}.
\end{equation}
The recursive evaluation gives the finite-step functional
\begin{equation}
\mcG_{\mathrm{perc}}^{\mathrm{CCM}}(\zeta,Q,\Gamma)
=
X_0(0)+\alpha Y_0(0)
-
\frac12\sum_{\ell=0}^R\zeta_\ell
\left(
Q_{\ell+1}\Gamma_{\ell+1}-Q_\ell\Gamma_\ell
\right).
\label{eq:intro-main-functional}
\end{equation}
For the replica-symmetric profile we take
\begin{equation}
Q_0=0,
\qquad
Q_1=Q,
\qquad
Q_2=1,
\end{equation}
\begin{equation}
\Gamma_0=0,
\qquad
\Gamma_1=\Gamma_2=\Gamma,
\end{equation}
with \(\zeta_0=0\) and \(\zeta_1=1\), the functional becomes
\begin{equation}
\mcG_{\mathrm{RS}}(Q,\Gamma)
=
\E_Z\log2\cosh(\sqrt\Gamma Z)
+
\alpha\E_Z\log\E_{Z'}
U(\sqrt QZ+\sqrt{1-Q}Z')
-
\frac12\Gamma(1-Q).
\label{eq:intro-rs-gardner}
\end{equation}
With the a posteriori identification \(Q=q\) and \(\Gamma=\widehat q\) \cite{SimplifiedParisiIII,FranchiniRSBWithoutReplicas}, this is Gardner's replica-symmetric formula for the generalized Ising perceptron \cite{GardnerSpace,GardnerDerridaOptimal,BolthausenNakajimaSunXu}.

The hard binary perceptron is obtained from the hard channel
\begin{equation}
U_\kappa(u)=\one\{u\ge\kappa\}.
\end{equation}
We specialize the constraint terminal to this channel; the replica-symmetric hard formula below uses a spin variance strictly below one. If
\begin{equation}
H(a)=\int_a^\infty \Dz,
\end{equation}
then the replica-symmetric hard specialization is
\begin{equation}
\mcG_{\mathrm{KM}}^{\mathrm{RS}}(Q,\Gamma;\alpha,\kappa)
=
\E_Z\log2\cosh(\sqrt\Gamma Z)
+
\alpha\E_Z\log
H\left(
\frac{\kappa-\sqrt QZ}{\sqrt{1-Q}}
\right)
-
\frac12\Gamma(1-Q).
\label{eq:intro-km}
\end{equation}
This is the Gardner--Krauth--M\'ezard expression for the one-sided binary perceptron \cite{KrauthMezardStorage,MezardGardnerCavity,AubinPerkinsZdeborovaSymmetric}.

\section{Positive auxiliary representation}

We start from the Ising perceptron with Gaussian patterns
\begin{equation}
Z_N
=
\sum_{\sigma\in\{-1,+1\}^N}
\prod_{\mu=1}^{M_N}
U\bigl(S_\mu(\sigma)\bigr),
\qquad
M_N=\alpha N+o(N),
\label{eq:perc-partition}
\end{equation}
where the field is defined as follows:
\begin{equation}
S_\mu(\sigma)
=
\frac1{\sqrt N}
\sum_{i=1}^{N}\xi_i^\mu\sigma_i,
\label{eq:perc-pattern-field}
\end{equation}
and the variables \(\xi_i^\mu\) are i.i.d. standard Gaussians.  The channel \(U\) is assumed nonnegative.  In the positive-channel calculation we assume that \(U\) has an exponential mixture representation
\begin{equation}
U(x)
=
\int_{\R} e^{tx}\,\rho(\dd t),
\qquad
\rho\ge 0.
\label{eq:positive-channel}
\end{equation}
Notice that the measure \(\rho\) need not be normalized.  If \(\rho\) is multiplied by a constant, then \(U\) is multiplied by the same constant and the pressure is shifted by the corresponding additive term.  Thus constants in the channel are kept inside \(U\).

For the positive auxiliary calculation we assume
\begin{equation}
0<\rho(\R)<\infty,
\qquad
\int_{\R} e^{(1/2+\eps)t^2}\rho(\dd t)<\infty
\quad\text{for some }\eps>0.
\label{eq:auxiliary-integrability}
\end{equation}
This condition makes the Gaussian channel convolutions and recursive averages used below finite and controls the auxiliary quadratic mass in the core comparison. It is satisfied by elementary exponential mixtures such as
\begin{equation}
U(x)=e^{hx},
\qquad
\rho=\delta_h,
\end{equation}
\begin{equation}
U(x)=2\cosh(hx),
\qquad
\rho=\delta_h+\delta_{-h},
\end{equation}
as well as by finite positive mixtures of exponentials with suitable growth.  The hard channel
\begin{equation}
U_\kappa(x)=\one\{x\ge\kappa\}
\end{equation}
is treated in Section~\ref{sec:hard-km} by specializing the constraint terminal of the recursion.

Substituting \eqref{eq:positive-channel} into \eqref{eq:perc-partition} gives
\begin{align}
Z_N
&=
\sum_{\sigma\in\{-1,+1\}^N}
\prod_{\mu=1}^{M_N}
\int_{\R} e^{t_\mu S_\mu(\sigma)}\rho(\dd t_\mu)
\nonumber\\
&=
\sum_{\sigma\in\{-1,+1\}^N}
\int_{\R^{M_N}}
\prod_{\mu=1}^{M_N}\rho(\dd t_\mu)
\exp\left\{
\frac1{\sqrt N}
\sum_{i=1}^{N}
\sum_{\mu=1}^{M_N}
\xi_i^\mu\sigma_i t_\mu
\right\}.
\label{eq:auxiliary-partition}
\end{align}
Thus the positive-channel perceptron is rewritten as a dense bipartite Gaussian model.  The two layers are denoted by
\begin{equation}
L_\sigma=\{1,\ldots,N\},
\qquad
L_t=\{1,\ldots,M_N\}.
\end{equation}
The variables on \(L_\sigma\) are Ising spins,
\begin{equation}
\sigma_i\in\{-1,+1\},
\end{equation}
while the variables on \(L_t\) are continuous auxiliary variables with base measure \(\rho\):
\begin{equation}
t_\mu\in\R,
\qquad
\rho(\dd t_\mu).
\end{equation}
The interaction exponent is
\begin{equation}
\mcH_N(\sigma,t)
=
\frac1{\sqrt N}
\sum_{i=1}^{N}
\sum_{\mu=1}^{M_N}
\xi_i^\mu\sigma_i t_\mu.
\label{eq:auxiliary-exponent}
\end{equation}
Equivalently, we can rewrite the partition function in the compact form
\begin{equation}
Z_N
=
\int e^{\mcH_N(\sigma,t)}\,\Pi_N(\dd\sigma,\dd t),
\label{eq:auxiliary-gibbs-form}
\end{equation}
where the reference measure is as follows:
\begin{equation}
\Pi_N(\dd\sigma,\dd t)
=
\sum_{\tau\in\{-1,+1\}^N}
\delta_\tau(\dd\sigma)
\prod_{\mu=1}^{M_N}\rho(\dd t_\mu).
\end{equation}
This is the form used by the CCM calculation.  The nonlinear channel factor has been replaced by a bilinear Gaussian interaction between the Ising layer and the auxiliary layer.  After the cavity step for a new constraint, the auxiliary variable is integrated out again and the original factor \(U\) reappears.

For a general channel, the spin-only representation need not be interface-only.  Let a blanket increment add new spins \(V_\ell\) to a past set \(Q_{\ell-1}\).  For a fixed pattern,
\begin{equation}
S_\mu(\sigma_{Q_\ell})
=
S_\mu(\sigma_{Q_{\ell-1}})+\Delta_\ell S_\mu,
\qquad
\Delta_\ell S_\mu
=
\frac1{\sqrt N}
\sum_{i\in V_\ell}\xi_i^\mu\sigma_i.
\end{equation}
The corresponding increment of the spin-only weight contains
\begin{equation}
\log U\bigl(S_\mu(\sigma_{Q_{\ell-1}})+\Delta_\ell S_\mu\bigr)
-
\log U\bigl(S_\mu(\sigma_{Q_{\ell-1}})\bigr).
\end{equation}
For a general channel, the logarithmic increment need not be a one-body term in the new spins.

In the auxiliary representation, by contrast, the exponent \eqref{eq:auxiliary-exponent} is bilinear.  If the old auxiliary variables are fixed, a new spin sees the Gaussian field
\begin{equation}
\frac1{\sqrt N}\sum_{\mu\in L_t}\xi_i^\mu t_\mu.
\end{equation}
If the old Ising variables are fixed, a new auxiliary variable sees the Gaussian field
\begin{equation}
\frac1{\sqrt N}\sum_{i\in L_\sigma}\xi_i^\mu\sigma_i.
\end{equation}
Thus, once the past is fixed, new variables in one layer enter linearly through fields generated by old variables in the other layer.  This is the interface structure used in the CCM geometry of the next section.

The auxiliary representation therefore has two roles.  First, it restores the bipartite edge structure required by the blanket decomposition.  Second, it separates the two cavity operations: adding an Ising spin gives a \(2\cosh\) factor, while adding a new perceptron constraint gives back the channel factor \(U\).  These two operations will become the site and constraint terms of the perceptron cavity functional.

\section{CCM geometry of the auxiliary bipartite model}

The positive auxiliary representation turns the perceptron into a dense bipartite Gaussian model with two layers,
\begin{equation}
L_\sigma=\{1,\ldots,N\},
\qquad
L_t=\{1,\ldots,M_N\},
\qquad
M_N=\alpha N+o(N).
\end{equation}
The variables on the first layer are Ising spins and the variables on the second layer are continuous auxiliary variables.  The interaction exponent is
\begin{equation}
\mcH_N(\sigma,t)
=
\frac1{\sqrt N}
\sum_{i\in L_\sigma}
\sum_{\mu\in L_t}
\xi_i^\mu\sigma_i t_\mu.
\label{eq:geometry-aux-exponent}
\end{equation}
This is the two-layer analogue of the RBM/DBM geometry used in the CCM.  A global blanket cuts the architectural layers, each cell is the intersection of one blanket with one layer, and the couplings are organized by cumulative rectangles and edge shells \cite{FranchiniConcettiBardellaDBM}.  In the present model there is only one edge family, between the Ising layer and the auxiliary layer.

We fix the blanket partition independently of the couplings. A blanket increment is written as
\begin{equation}
V_\ell=V_\ell^\sigma\sqcup V_\ell^t,
\qquad
V_\ell^\sigma=V_\ell\cap L_\sigma,
\qquad
V_\ell^t=V_\ell\cap L_t.
\end{equation}
In the following we use the convention that
\begin{equation}
U_\ell^\sigma=\bigcup_{s<\ell}V_s^\sigma,
\qquad
U_\ell^t=\bigcup_{s<\ell}V_s^t
\end{equation}
is the past before adding the \(\ell\)-th blanket increment.  This is only an index convention.  In the cumulative-after notation used in the RBM/DBM geometry, the same set would be denoted by the previous cumulative blanket.  With the present convention, the shell added at level \(\ell\) is \(B_{\ell+1}\setminus B_\ell\).  The cumulative rectangle before adding level \(\ell\) is
\begin{equation}
B_\ell=U_\ell^\sigma\times U_\ell^t.
\end{equation}
After adding \(V_\ell^\sigma\) and \(V_\ell^t\), the next rectangle is
\begin{equation}
B_{\ell+1}
=
(U_\ell^\sigma\cup V_\ell^\sigma)
\times
(U_\ell^t\cup V_\ell^t).
\end{equation}
The edge shell of level \(\ell\) is therefore 
\begin{equation}
W_\ell=B_{\ell+1}\setminus B_\ell.
\end{equation}
Expanding the rectangle gives the disjoint decomposition
\begin{equation}
W_\ell
=
(V_\ell^\sigma\times U_\ell^t)
\cup
(U_\ell^\sigma\times V_\ell^t)
\cup
(V_\ell^\sigma\times V_\ell^t).
\label{eq:geometry-shell-decomposition}
\end{equation}
The first two pieces form the interface,
\begin{equation}
W_\ell^{\mathrm{int}}
=
(V_\ell^\sigma\times U_\ell^t)
\cup
(U_\ell^\sigma\times V_\ell^t),
\end{equation}
and the last piece is the core, explicitly
\begin{equation}
W_\ell^{\mathrm{core}}
=
V_\ell^\sigma\times V_\ell^t.
\end{equation}
Accordingly, the layer increment of the exponent splits as
\begin{equation}
\mcH_\ell
=
\mcH_\ell^{\mathrm{int}}
+
\mcH_\ell^{\mathrm{core}}.
\label{eq:geometry-interface-core-exponent}
\end{equation}
The core is denoted only by \(\mcH_\ell^{\mathrm{core}}\) throughout the paper.

The spin side is described by a constructive variance profile
\begin{equation}
0=Q_0\le Q_1\le\cdots\le Q_{R+1}=1.
\end{equation}
At finite volume, \(Q_\ell\) is represented by the normalized amount of spin-side blanket that has already been included.  The auxiliary side is described by a second constructive variance profile
\begin{equation}
0=\Gamma_0\le\Gamma_1\le\cdots\le\Gamma_R<\infty,
\qquad
\Gamma_{R+1}:=\Gamma_R.
\end{equation}
The auxiliary layer is continuous, so its natural coordinate is a quadratic variance scale.  In the finite volume setting the relevant quantity is
\begin{equation}
\frac1N\sum_{\mu\in U_\ell^t}t_\mu^2.
\end{equation}
In the constructive functional this scale is represented by \(\Gamma_\ell\).  The profiles \(Q_\ell\) and \(\Gamma_\ell\) are constructive variance profiles.

The cumulative bipartite scale up to level \(\ell\) is
\begin{equation}
Q_\ell\Gamma_\ell.
\end{equation}
Thus the shell scale at level \(\ell\) is
\begin{equation}
\Delta(Q\Gamma)_\ell
=
Q_{\ell+1}\Gamma_{\ell+1}-Q_\ell\Gamma_\ell.
\end{equation}
Recalling the notation for the difference
\begin{equation}
\Delta Q_\ell=Q_{\ell+1}-Q_\ell,
\qquad
\Delta\Gamma_\ell=\Gamma_{\ell+1}-\Gamma_\ell,
\end{equation}
one has the following expression:
\begin{equation}
\Delta(Q\Gamma)_\ell
=
Q_\ell\Delta\Gamma_\ell
+
\Gamma_\ell\Delta Q_\ell
+
\Delta Q_\ell\Delta\Gamma_\ell.
\label{eq:geometry-product-increment}
\end{equation}
The first two terms are the interface scale.  The last term is the core scale.  The total core scale vanishes under refinement,
\begin{equation}
\sum_{\ell=0}^{R}
\Delta Q_\ell\Delta\Gamma_\ell\longrightarrow0.
\label{eq:geometry-refinement}
\end{equation}
The pressure effect of this refinement is controlled by \eqref{eq:core-pressure-bound} below.
The interface fields are one-body after conditioning on the past.  A new spin \(i\in V_\ell^\sigma\) sees the old auxiliary side through
\begin{equation}
h_i^\sigma
=
\frac1{\sqrt N}
\sum_{\mu\in U_\ell^t}
\xi_i^\mu t_\mu.
\label{eq:geometry-spin-field}
\end{equation}
Conditioned on the old auxiliary variables, this is a centered Gaussian with variance
\begin{equation}
\operatorname{Var}(h_i^\sigma\mid t)
=
\frac1N\sum_{\mu\in U_\ell^t}t_\mu^2,
\end{equation}
represented constructively by \(\Gamma_\ell\).  A new auxiliary variable \(\mu\in V_\ell^t\) sees the old Ising side through
\begin{equation}
h_\mu^t
=
\frac1{\sqrt N}
\sum_{i\in U_\ell^\sigma}
\xi_i^\mu\sigma_i.
\label{eq:geometry-aux-field}
\end{equation}
Conditioned on the old spin configuration, this is a centered Gaussian with variance
\begin{equation}
\operatorname{Var}(h_\mu^t\mid\sigma)
=
\frac{|U_\ell^\sigma|}{N},
\end{equation}
represented constructively by \(Q_\ell\).  Hence the interface exponent has the one-body form
\begin{equation}
\mcH_\ell^{\mathrm{int}}
=
\sum_{i\in V_\ell^\sigma}h_i^\sigma\sigma_i
+
\sum_{\mu\in V_\ell^t}h_\mu^t t_\mu.
\label{eq:geometry-interface-one-body}
\end{equation}
This is the basic interface property needed for the CCM calculation.

The core exponent (core Hamiltonian) is explicitly
\begin{equation}
\mcH_\ell^{\mathrm{core}}
=
\frac1{\sqrt N}
\sum_{i\in V_\ell^\sigma}
\sum_{\mu\in V_\ell^t}
\xi_i^\mu\sigma_i t_\mu.
\label{eq:geometry-core-exponent}
\end{equation}
Conditioned on the new auxiliary variables, its variance is
\begin{equation}
\operatorname{Var}(\mcH_\ell^{\mathrm{core}}\mid t)
=
\frac1N
|V_\ell^\sigma|
\sum_{\mu\in V_\ell^t}t_\mu^2.
\end{equation}
At the constructive scale this is of order
\begin{equation}
N\Delta Q_\ell\Delta\Gamma_\ell.
\end{equation}
Write \(\overline Z_N\) for the partition function with all core couplings removed. Gaussian core comparison, as in the RBM/DBM calculation~\cite{FranchiniConcettiBardellaDBM}, with the auxiliary quadratic mass controlled by \eqref{eq:auxiliary-integrability}, gives
\begin{equation}
0\leq\frac1N\E\log\frac{Z_N}{\overline Z_N}
\leq
\frac{M_N}{2N\eps}
\max_\ell\frac{|V_\ell^\sigma|}{N}
\log\!\left[
\frac{\int_{\R}e^{(1/2+\eps)t^2}\rho(\dd t)}{\rho(\R)}
\right].
\label{eq:core-pressure-bound}
\end{equation}
The right-hand side vanishes when the largest spin-side blanket fraction tends to zero. We first take the thermodynamic limit at a fixed partition, then refine it. This gives the interface-only pressure, while the correction field retains the complete finite shell increments.
We synthesize the meaning of these variables into the following geometric dictionary:
\begin{equation}
\text{new spin field}\,\leftrightarrow\, \Gamma, \quad
\text{new constraint field}\,\leftrightarrow\, Q, \quad
\text{correction field}\,\leftrightarrow\, Q\Gamma. \quad
\end{equation}
The next section writes the scalar cavity functional whose three terms are read through these constructive scales.

\section{The perceptron cavity functional}

We now write the scalar cavity functional that will be evaluated by the constructive calculation. The functional is the perceptron analogue of the Aizenman--Sims--Starr cavity increment \cite{AizenmanSimsStarr}. It contains one term for adding an Ising spin, one term for adding perceptron constraints at density \(\alpha\), and one cavity correction term. The only input from the channel is the scalar function \(U\).

Let \(\nu\) be a trial reservoir supported on configurations of the auxiliary bipartite model. We write its atoms as follows:
\begin{equation}
a\in\mathcal I,
\qquad
w_a\ge0,
\qquad
\sum_{a\in\mathcal I}w_a=1.
\end{equation}
The atom \(a\) carries an Ising configuration and an auxiliary configuration,
\begin{equation}
\sigma^a=(\sigma_i^a)_{1\le i\le N},
\qquad
t^a=(t_\mu^a)_{1\le \mu\le M_N}.
\end{equation}
For an observable \(F^a\), we use the following braket notation for the average
\begin{equation}
\langle F\rangle_\nu
=
\sum_{a\in\mathcal I}w_aF^a.
\end{equation}
Reservoir atoms are denoted by \(a,b\). The density of constraints is always denoted by \(\alpha\).

\subsection{Adding an Ising spin}

The first cavity operation adds one Ising spin. The new spin interacts with the old auxiliary variables through fresh Gaussian couplings. For the reservoir atom \(a\), define
\begin{equation}
x^a
=
\frac1{\sqrt N}
\sum_{\mu=1}^{M_N}
\widetilde\xi_\mu t_\mu^a,
\label{eq:ass-new-spin-field}
\end{equation}
where \(\{\widetilde\xi_\mu\}\) is independent of the reservoir and of all previous disorder.

If the new spin is \(\sigma_0\in\{-1,+1\}\), its Boltzmann factor is
\begin{equation}
\exp\{\sigma_0x^a\}.
\end{equation}
Summing over \(\sigma_0\) gives the usual hyperbolic factor
\begin{equation}
\sum_{\sigma_0=\pm1}e^{\sigma_0x^a}
=
2\cosh x^a.
\end{equation}
Thus the new-spin contribution to the scalar cavity increment is
\begin{equation}
\log\langle2\cosh x\rangle_\nu.
\end{equation}
Conditioned on the reservoir, \(x^a\) is a centered Gaussian process with covariance
\begin{equation}
\E x^a x^b
=
\frac1N
\sum_{\mu=1}^{M_N}
t_\mu^a t_\mu^b.
\label{eq:ass-x-covariance}
\end{equation}
In the constructive calculation this empirical covariance scale is read through the auxiliary-side variance profile \(\Gamma\).

\subsection{Adding perceptron constraints}

The second cavity operation adds new constraints. Since \(M_N=\alpha N+o(N)\), the number of constraints grows at density \(\alpha\) when the number of spins grows. A fresh constraint sees the old Ising configuration through the field
\begin{equation}
s^a
=
\frac1{\sqrt N}
\sum_{i=1}^{N}
\breve\xi_i\sigma_i^a,
\label{eq:ass-new-constraint-field}
\end{equation}
where \(\{\breve\xi_i\}\) is another fresh independent Gaussian family.
In the positive auxiliary representation, the auxiliary variable of the fresh constraint is integrated out:
\begin{equation}
\int e^{ts^a}\rho(\dd t)
=
U(s^a).
\end{equation}
Thus one fresh constraint contributes \(U(s^a)\). Since fresh constraints are added with density \(\alpha\), the constraint contribution to the cavity functional is
\begin{equation}
\alpha\log\langle U(s)\rangle_\nu.
\end{equation}
Conditioned on the reservoir, \(s^a\) is centered Gaussian with covariance
\begin{equation}
\E s^a s^b
=
\frac1N
\sum_{i=1}^{N}
\sigma_i^a\sigma_i^b.
\label{eq:ass-s-covariance}
\end{equation}
In the constructive calculation this empirical covariance scale is read through the Ising-side variance profile \(Q\).

\subsection{The cavity correction}

The third term is the cavity correction. It is the bipartite analogue of the usual Aizenman--Sims--Starr correction term. When the system size is changed, the old dense interaction is rescaled. The correction is represented by a fresh Gaussian process
\begin{equation}
y^a
=
\frac1N
\sum_{i=1}^{N}
\sum_{\mu=1}^{M_N}
\overline\xi_i^\mu\sigma_i^a t_\mu^a,
\label{eq:ass-correction-field}
\end{equation}
where \(\{\overline\xi_i^\mu\}\) is independent of the Gaussian families used in \(x\) and \(s\), and independent of the reservoir. The factor \(1/N\) is chosen so that the covariance of \(y\) is the product of the spin-side and auxiliary-side covariance scales.

Conditioned on the reservoir, \(y^a\) is centered Gaussian with covariance
\begin{equation}
\E y^a y^b
=
\left(
\frac1N
\sum_{i=1}^{N}
\sigma_i^a\sigma_i^b
\right)
\left(
\frac1N
\sum_{\mu=1}^{M_N}
t_\mu^a t_\mu^b
\right).
\label{eq:ass-y-covariance}
\end{equation}
The constructive scale of this covariance is the product \(Q\Gamma\). Notice that the correction term enters the cavity increment with a negative sign,
\begin{equation}
-\log\langle e^y\rangle_\nu.
\end{equation}

\subsection{The scalar functional}

Combining the three cavity operations gives
\begin{equation}
\mcA_{\mathrm{perc}}(\nu)
=
\E\left[
\log\langle2\cosh x\rangle_\nu
+
\alpha\log\langle U(s)\rangle_\nu
-
\log\langle e^y\rangle_\nu
\right].
\label{eq:ass-perceptron-functional}
\end{equation}
Equivalently,
\begin{equation}
\mcA_{\mathrm{perc}}(\nu)
=
\E\left[
\log\sum_{a\in\mathcal I}w_a\,2\cosh x^a
+
\alpha\log\sum_{a\in\mathcal I}w_a\,U(s^a)
-
\log\sum_{a\in\mathcal I}w_a\,e^{y^a}
\right].
\end{equation}
The expectation is over the fresh Gaussian families defining \(x,s,y\), and over the reservoir randomness if \(\nu\) is random.

The blanket decomposition expresses the three covariance scales in \eqref{eq:ass-x-covariance}, \eqref{eq:ass-s-covariance}, and \eqref{eq:ass-y-covariance} through the constructive profiles
\begin{equation}
\Gamma,
\qquad
Q,
\qquad
Q\Gamma,
\end{equation}
respectively. This gives the dictionary
\begin{equation}
x\longleftrightarrow \Gamma,
\qquad
s\longleftrightarrow Q,
\qquad
y\longleftrightarrow Q\Gamma.
\end{equation}
The next section constructs the corresponding cavity variables and identifies them with the ROSt variables.

\section{Constructive cavity marks}\label{sec:constructive-marks}

We now write the blanket components of the three cavity variables using the scales computed above.  The Ising side carries the profile \(Q\), the auxiliary side carries the profile \(\Gamma\), and the bipartite correction carries the product scale \(Q\Gamma\).
We use the finite profiles
\begin{equation}
0=Q_0\le Q_1\le\cdots\le Q_{R+1}=1,
\end{equation}
\begin{equation}
0=\Gamma_0\le \Gamma_1\le\cdots\le \Gamma_R<\infty,
\qquad
\Gamma_{R+1}:=\Gamma_R.
\end{equation}
At a fixed variance profile, a final auxiliary variance increment contributes half its variance to both the site contribution and the cavity correction after the final thermal average, since the terminal spin variance is one. These contributions cancel in the cavity functional, which justifies the convention above. The following cavity marks incorporate this reduction, while the constraint field retains its final residual variance.
We write
\begin{equation}
\Delta Q_\ell=Q_{\ell+1}-Q_\ell,
\qquad
\Delta\Gamma_\ell=\Gamma_{\ell+1}-\Gamma_\ell,
\end{equation}
such that the shell increment is written as follows:
\begin{equation}
\Delta(Q\Gamma)_\ell
=
Q_{\ell+1}\Gamma_{\ell+1}-Q_\ell\Gamma_\ell,
\qquad
\ell=0,\ldots,R.
\end{equation}
All three quantities are non-negative under the monotonicity assumptions on \(Q\) and \(\Gamma\).
We use hierarchical labels and their prefixes,
\begin{equation}
\mathbf a=(a_0,\ldots,a_R),
\qquad
\mathbf a_{\le\ell}=(a_0,\ldots,a_\ell).
\end{equation}
Let \(w_{\mathbf a}\) be the normalized Ruelle-cascade weights of the ROSt associated with \(\zeta\)~\cite{RuelleREM,BolthausenSznitmanCascades,ArguinCascades},
\begin{equation}
w_{\mathbf a}\ge0,
\qquad
\sum_{\mathbf a}w_{\mathbf a}=1.
\end{equation}
The root label \(a_0\) is fixed and its Gaussian variables are common to all states. The cascade weights concern the intermediate labels; summation over \(a_R\) denotes the final Gaussian average inside each bracket. Section~\ref{sec:recursive-evaluation} implements these endpoint conventions. We use the letter \(w\) for the weights, while \(\xi_i^\mu\) is reserved for the pattern disorder.
At each node introduce three independent standard Gaussian variables,
\begin{equation}
z_\ell(\mathbf a_{\le\ell}),
\qquad
\eta_\ell(\mathbf a_{\le\ell}),
\qquad
g_\ell(\mathbf a_{\le\ell}).
\end{equation}
The families are independent of the weights and at distinct nodes. States with the same prefix use the same Gaussian variable. The density of constraints remains denoted by \(\alpha\).

The new-spin field is generated by the old auxiliary side.  Its constructive variance increments are therefore the increments of \(\Gamma\).  The CCM mark for the new-spin field is
\begin{equation}
x^{\mathbf a}
=
\sum_{\ell=0}^{R}
\sqrt{\Delta\Gamma_\ell}\,
z_\ell(\mathbf a_{\le\ell}).
\label{eq:constructive-x}
\end{equation}
The terminal factor associated with the new Ising spin is
\begin{equation}
2\cosh x^{\mathbf a}.
\end{equation}
This is the constructive reading of the physical field
\begin{equation}
x^a=
\frac1{\sqrt N}
\sum_\mu \widetilde\xi_\mu t_\mu^a,
\end{equation}
whose conditional variance is determined by the quadratic mass of the auxiliary side.

The new-constraint field is generated by the old Ising side.  Its constructive variance increments are the increments of \(Q\).  The CCM mark for the new-constraint field is
\begin{equation}
s^{\mathbf a}
=
\sum_{\ell=0}^{R}
\sqrt{\Delta Q_\ell}\,
\eta_\ell(\mathbf a_{\le\ell}).
\label{eq:constructive-s}
\end{equation}
The terminal factor associated with a new perceptron constraint is
\begin{equation}
U(s^{\mathbf a}).
\end{equation}
The final increment is the residual Gaussian variance inside the constraint channel:
\begin{equation}
\Delta Q_R=1-Q_R
\end{equation}
This residual variance is the source of the Gaussian smoothing in the hard-channel specialization.

The correction field is bipartite.  Its cumulative scale is the product of the two constructive profiles, \(Q_\ell\Gamma_\ell\). Accordingly the correction mark is
\begin{equation}
y^{\mathbf a}
=
\sum_{\ell=0}^{R}
\sqrt{\Delta(Q\Gamma)_\ell}\,
g_\ell(\mathbf a_{\le\ell}).
\label{eq:constructive-y}
\end{equation}
Its terminal factor is
\(e^{y^{\mathbf a}}\).
This is the constructive reading of the physical correction field
\begin{equation}
y^a
=
\frac1N
\sum_{i,\mu}
\overline\xi_i^\mu\sigma_i^a t_\mu^a,
\end{equation}
whose covariance scale is the product of the Ising-side and auxiliary-side scales.

CCM identifies the constructive variables \eqref{eq:constructive-x}, \eqref{eq:constructive-s}, and \eqref{eq:constructive-y} with the Aizenman--Sims--Starr ROSt variables~\cite{SimplifiedParisiIII}. The constructive perceptron cavity functional is therefore
\begin{equation}
\mcA_{\mathrm{perc}}^{\mathrm{constr}}(w)
=
\E\left[
\log\sum_{\mathbf a}w_{\mathbf a}\,2\cosh x^{\mathbf a}
+
\alpha\log\sum_{\mathbf a}w_{\mathbf a}\,U(s^{\mathbf a})
-
\log\sum_{\mathbf a}w_{\mathbf a}\,e^{y^{\mathbf a}}
\right].
\label{eq:constructive-functional-sums}
\end{equation}
Equivalently, using the braket notation:
\begin{equation}
\mcA_{\mathrm{perc}}^{\mathrm{constr}}(w)
=
\E\left[
\log\langle2\cosh x\rangle_w
+
\alpha\log\langle U(s)\rangle_w
-
\log\langle e^y\rangle_w
\right].
\label{eq:constructive-functional-brackets}
\end{equation}
Here \(\langle\cdot\rangle_w\) is the normalized ROSt average. The expectation is over the Gaussian variables and the ROSt weights.

The dictionary of the construction is
\begin{equation}
x\leftrightarrow\Gamma,
\qquad
s\leftrightarrow Q,
\qquad
y\leftrightarrow Q\Gamma.
\end{equation}
The profiles \(Q\) and \(\Gamma\) are constructive variance profiles.  The comparison with the usual perceptron notation is made later, after the CCM computation has produced the functional.

\section{Recursive evaluation}\label{sec:recursive-evaluation}

We evaluate the three normalized brackets in \eqref{eq:constructive-functional-brackets} by Parisi's recursive averaging~\cite{SimplifiedParisiIII,RuelleREM,BolthausenSznitmanCascades,ArguinCascades}. Each level transforms a terminal function by a logarithmic Gaussian convolution.
For a function \(f\), a variance increment \(a\ge0\), and \(0<m\le1\), define
\begin{equation}
T_{m,a}f(u)
=
\frac1m
\log
\E_Z\exp\{m f(u+\sqrt a Z)\}.
\label{eq:ppp-transform}
\end{equation}
For \(m=0\) we use the limit expression
\begin{equation}
T_{0,a}f(u)
=
\E_Z f(u+\sqrt a Z).
\end{equation}
These transforms are applied backwards from level \(R\) to level \(0\), with the earlier Gaussian variables held fixed. The endpoint \(\zeta_0=0\) gives the ordinary average over the common field, while \(\zeta_R=1\) gives the terminal thermal average. Zero-variance steps act as the identity, and consecutive equal exponents merge by Gaussian convolution.
For the site term, set the following starting point
\begin{equation}
X_{R+1}(u)=\log2\cosh u.
\end{equation}
Since the site mark \(x\) has variance increments \(\Delta\Gamma_\ell\), define
\begin{equation}
X_\ell
=
T_{\zeta_\ell,\Delta\Gamma_\ell}X_{\ell+1},
\qquad
\ell=R,R-1,\ldots,0.
\label{eq:site-recursion}
\end{equation}
In explicit form we can alternatively write:
\begin{equation}
X_\ell(u)
=
\frac1{\zeta_\ell}
\log
\E_Z
\exp\left\{
\zeta_\ell
X_{\ell+1}(u+\sqrt{\Delta\Gamma_\ell}Z)
\right\},
\end{equation}
with the expectation convention when \(\zeta_\ell=0\).  Therefore
\begin{equation}
\E\log\langle2\cosh x\rangle_w
=
X_0(0).
\label{eq:site-term-result}
\end{equation}
For the constraint term, set the starting point
\begin{equation}
Y_{R+1}(u)=\log U(u).
\end{equation}
Since the constraint mark \(s\) has variance increments \(\Delta Q_\ell\), define
\begin{equation}
Y_\ell
=
T_{\zeta_\ell,\Delta Q_\ell}Y_{\ell+1},
\qquad
\ell=R,R-1,\ldots,0.
\label{eq:constraint-recursion}
\end{equation}
Explicitly, this reads:
\begin{equation}
Y_\ell(u)
=
\frac1{\zeta_\ell}
\log
\E_Z
\exp\left\{
\zeta_\ell
Y_{\ell+1}(u+\sqrt{\Delta Q_\ell}Z)
\right\},
\end{equation}
with the expectation convention when \(\zeta_\ell=0\).  Hence
\begin{equation}
\E\log\langle U(s)\rangle_w
=
Y_0(0).
\label{eq:constraint-term-result}
\end{equation}
The last increment \(\Delta Q_R=1-Q_R\) remains in this recursion.  It is the residual variance of the perceptron constraint field.
For the correction term, set
\begin{equation}
C_{R+1}(u)=u.
\end{equation}
The correction mark has variance increments \(\Delta(Q\Gamma)_\ell\), hence
\begin{equation}
C_\ell=T_{\zeta_\ell,\Delta(Q\Gamma)_\ell}C_{\ell+1}.
\end{equation}
If \(C_{\ell+1}(u)=u+c_{\ell+1}\), then
\begin{align}
T_{\zeta_\ell,a}C_{\ell+1}(u)
&=
\frac1{\zeta_\ell}
\log
\E_Z
\exp\{\zeta_\ell(u+\sqrt a Z+c_{\ell+1})\}  \\
&=
 u+c_{\ell+1}+\frac12\zeta_\ell a,
\end{align}
for \(\zeta_\ell>0\), while for \(\zeta_\ell=0\) the added contribution is zero.  Iterating gives
\begin{equation}
C_0(0)
=
\frac12
\sum_{\ell=0}^{R}
\zeta_\ell\Delta(Q\Gamma)_\ell.
\end{equation}
Thus, the fugacity correction is
\begin{equation}
\E\log\langle e^y\rangle_w
=
\frac12
\sum_{\ell=0}^{R}
\zeta_\ell
\left(
Q_{\ell+1}\Gamma_{\ell+1}-Q_\ell\Gamma_\ell
\right).
\label{eq:correction-term-result}
\end{equation}
Substituting \eqref{eq:site-term-result}, \eqref{eq:constraint-term-result}, and \eqref{eq:correction-term-result} into the constructive cavity functional gives
\begin{equation}
\mcG_{\mathrm{perc}}^{\mathrm{CCM}}
(\zeta,Q,\Gamma)
=
X_0(0)
+
\alpha Y_0(0)
-
\frac12
\sum_{\ell=0}^{R}
\zeta_\ell
\left(
Q_{\ell+1}\Gamma_{\ell+1}-Q_\ell\Gamma_\ell
\right).
\label{eq:ppp-final-functional}
\end{equation}
The first term is the Ising site contribution, the second is the perceptron constraint contribution at density \(\alpha\), and the last term is the bipartite cavity correction.

The computation separates the two channels: \(X\) depends only on the auxiliary profile \(\Gamma\), while \(Y\) depends only on the spin profile \(Q\).  The coupling between the two profiles appears only in the cavity correction through the cumulative product scale \(Q\Gamma\).

\section{Replica-symmetric specialization and Gardner formula}
\label{sec:rs-gardner}

We now specialize the finite-step CCM functional to the replica-symmetric profile and identify the resulting expression with Gardner's formula for the generalized Ising perceptron.

The finite-step functional is
\begin{equation}
\mcG_{\mathrm{perc}}^{\mathrm{CCM}}
=
X_0(0)
+
\alpha Y_0(0)
-
\frac12
\sum_{\ell=0}^{R}
\zeta_\ell
\left(
Q_{\ell+1}\Gamma_{\ell+1}-Q_\ell\Gamma_\ell
\right).
\label{eq:rs-start-functional-revised}
\end{equation}
The replica-symmetric specialization uses two variance steps on the spin side and one nontrivial step on the auxiliary side: we take
\begin{equation}
Q_0=0,
\qquad
Q_1=Q,
\qquad
Q_2=1,
\label{eq:rs-Q-profile}
\end{equation}
\begin{equation}
\Gamma_0=0,
\qquad
\Gamma_1=\Gamma_2=\Gamma,
\label{eq:rs-Gamma-profile}
\end{equation}
\begin{equation}
\zeta_0=0,
\qquad
\zeta_1=1.
\label{eq:rs-zeta-profile}
\end{equation}
Thus the constraint field has common component of variance \(Q\) and residual component of variance \(1-Q\).  The site field has variance \(\Gamma\), with no final residual increment.  This asymmetry is natural in the auxiliary perceptron: the spin side is bounded and contributes a \(2\cosh\) terminal, while the constraint side retains the final channel convolution.
The site recursion starts from
\begin{equation}
X_2(u)=\log 2\cosh u.
\end{equation}
Since \(\Gamma_2-\Gamma_1=0\), the top step is the identity.  The lower step has variance \(\Gamma\) and parameter \(\zeta_0=0\), hence it is an ordinary Gaussian expectation:
\begin{equation}
X_0(0)
=
\E_Z\log 2\cosh(\sqrt\Gamma Z).
\label{eq:rs-site-term-revised}
\end{equation}
For the constraint term, the terminal is
\begin{equation}
Y_2(u)=\log U(u).
\end{equation}
The upper step has variance \(1-Q\) and parameter \(\zeta_1=1\), so
\begin{equation}
Y_1(u)
=
\log \E_{Z'} U(u+\sqrt{1-Q}Z').
\label{eq:rs-Y1-revised}
\end{equation}
The lower step has variance \(Q\) and parameter \(\zeta_0=0\), therefore
\begin{equation}
Y_0(0)
=
\E_Z
\log \E_{Z'}
U(\sqrt Q Z+\sqrt{1-Q}Z').
\label{eq:rs-constraint-term-revised}
\end{equation}
The correction (or fugacity) term is
\begin{equation}
-
\frac12
\sum_{\ell=0}^{1}
\zeta_\ell
\left(
Q_{\ell+1}\Gamma_{\ell+1}-Q_\ell\Gamma_\ell
\right).
\end{equation}
The first term vanishes because \(\zeta_0=0\).  The second term is
\begin{equation}
Q_2\Gamma_2-Q_1\Gamma_1
=
\Gamma(1-Q),
\end{equation}
and hence the correction is rewritten again:
\begin{equation}
-
\frac12\Gamma(1-Q).
\label{eq:rs-correction-revised}
\end{equation}
Combining \eqref{eq:rs-site-term-revised}, \eqref{eq:rs-constraint-term-revised}, and \eqref{eq:rs-correction-revised}, we obtain
\begin{equation}
\mcG_{\mathrm{RS}}(Q,\Gamma)
=
\E_Z\log2\cosh(\sqrt\Gamma Z)
+
\alpha
\E_Z
\log
\E_{Z'}
U(\sqrt QZ+\sqrt{1-Q}Z')
-
\frac12\Gamma(1-Q).
\label{eq:rs-gardner-formula-revised}
\end{equation}
This is the Gardner replica-symmetric expression for the generalized Ising perceptron.  In the CCM derivation, \(Q\) and \(\Gamma\) are constructive variance parameters.  After the CCM calculation, the usual perceptron notation is recovered by setting
\begin{equation}
Q=q,
\qquad
\Gamma=\widehat q.
\label{eq:gardner-identification}
\end{equation}
With this identification, \eqref{eq:rs-gardner-formula-revised} coincides with the classical RS Gardner functional \cite{GardnerSpace,GardnerDerridaOptimal,MezardGardnerCavity,BolthausenNakajimaSunXu}.
It is useful to see the stationary equations, because they also match the standard form.  Define
\begin{equation}
L_Q(u)
=
\log \E_Z U(u+\sqrt{1-Q}Z),
\qquad
F_Q(u)=\partial_u L_Q(u).
\label{eq:LQ-FQ-def}
\end{equation}
Then the formula is rewritten as
\begin{equation}
\mcG_{\mathrm{RS}}(Q,\Gamma)
=
\E_Z\log2\cosh(\sqrt\Gamma Z)
+
\alpha\E_Z L_Q(\sqrt QZ)
-
\frac12\Gamma(1-Q).
\end{equation}
Differentiating with respect to \(\Gamma\), and using Gaussian integration by parts, gives
\begin{equation}
Q
=
\E_Z\tanh^2(\sqrt\Gamma Z).
\label{eq:rs-stationarity-Q-revised}
\end{equation}
Differentiating with respect to \(Q\) gives
\begin{equation}
\Gamma
=
-2\alpha
\frac{\partial}{\partial Q}
\E_Z
\log
\E_{Z'}U(\sqrt QZ+\sqrt{1-Q}Z').
\label{eq:rs-stationarity-Gamma-derivative}
\end{equation}
Under the regularity assumptions needed to differentiate the channel convolution, equivalently
\begin{equation}
\Gamma
=
\alpha\E_Z F_Q(\sqrt QZ)^2.
\label{eq:rs-stationarity-Gamma-F}
\end{equation}
The two equations \eqref{eq:rs-stationarity-Q-revised} and \eqref{eq:rs-stationarity-Gamma-F} are indeed the Gardner self-consistency equations, written in the CCM notation.
These stationarity equations characterize interior replica-symmetric candidates. Boundary candidates are tested by admissible one-sided variations. The soft variational selection is discussed in Section~\ref{sec:openai-comparison}; higher-step profiles are treated below by the same Parisi recursion.

\section{Hard-channel specialization and Krauth--M\'ezard formula}
\label{sec:hard-km}

We now specialize the constraint channel to the one-sided hard perceptron.  The hard channel is
\begin{equation}
U_\kappa(u)=\one\{u\ge \kappa\}.
\label{eq:hard-channel-def-revised}
\end{equation}
We specialize the constraint terminal of the CCM recursion to this channel. The final Gaussian convolution gives a finite logarithmic terminal whenever its residual variance is positive.
Let
\begin{equation}
H(a)=\int_a^\infty \Dz,
\qquad
\Dz=\frac{e^{-z^2/2}}{\sqrt{2\pi}}\,\dd z.
\label{eq:H-function-def}
\end{equation}
In the RS profile, the constraint field and its domain are
\begin{equation}
\sqrt QZ+\sqrt{1-Q}Z',
\qquad 0\leq Q<1.
\end{equation}
Therefore, we have the following:
\begin{equation}
\E_{Z'} U_\kappa(\sqrt QZ+\sqrt{1-Q}Z')
=
H\left(
\frac{\kappa-\sqrt QZ}{\sqrt{1-Q}}
\right).
\label{eq:hard-channel-convolution-revised}
\end{equation}
Substituting this expression into \eqref{eq:rs-gardner-formula-revised} gives
\begin{equation}
\mcG_{\mathrm{KM}}^{\mathrm{RS}}(Q,\Gamma;\alpha,\kappa)
=
\E_Z\log2\cosh(\sqrt\Gamma Z)
+
\alpha\E_Z\log
H\left(
\frac{\kappa-\sqrt QZ}{\sqrt{1-Q}}
\right)
-
\frac12\Gamma(1-Q).
\label{eq:km-rs-formula-revised}
\end{equation}
After the a posteriori identification
\begin{equation}
Q=q,
\qquad
\Gamma=\widehat q,
\end{equation}
this is the Gardner--Krauth--M\'ezard expression for the one-sided binary perceptron \cite{KrauthMezardStorage,MezardGardnerCavity,AubinPerkinsZdeborovaSymmetric}.  The correction term becomes in fact
\begin{equation}
-
\frac12\Gamma(1-Q)
=
\frac12(q-1)\widehat q,
\end{equation}
which is the usual sign convention in that literature.

The stationary equations follow from the soft RS equations by specializing the channel.  The equation from the site term remains
\begin{equation}
Q=\E_Z\tanh^2(\sqrt\Gamma Z).
\label{eq:km-stationarity-Q-revised}
\end{equation}
For the constraint side, define
\begin{equation}
A_Q(u)=\frac{\kappa-u}{\sqrt{1-Q}},
\qquad
\varphi(a)=\frac{e^{-a^2/2}}{\sqrt{2\pi}}.
\end{equation}
The derivative of the hard-channel logarithm is
\begin{equation}
\partial_u\log H(A_Q(u))
=
\frac{\varphi(A_Q(u))}{\sqrt{1-Q}\,H(A_Q(u))}.
\end{equation}
Hence \eqref{eq:rs-stationarity-Gamma-F} becomes the formula:
\begin{equation}
\Gamma
=
\alpha\E_Z
\left[
\frac{
\varphi\!\left(
\frac{\kappa-\sqrt QZ}{\sqrt{1-Q}}
\right)^2
}{
(1-Q)
H\!\left(
\frac{\kappa-\sqrt QZ}{\sqrt{1-Q}}
\right)^2
}
\right].
\label{eq:km-stationarity-Gamma-revised}
\end{equation}
Equations \eqref{eq:km-stationarity-Q-revised} and \eqref{eq:km-stationarity-Gamma-revised} are exactly the Gardner--Krauth--M\'ezard self-consistency equations in the variables \((Q,\Gamma)\).
For fixed \(\kappa\), the entropy on the selected RS stationary branch is
\begin{equation}
\Sigma_{\mathrm{KM}}(\alpha,\kappa)
=
\operatorname*{extr}_{Q,\Gamma}
\mcG_{\mathrm{KM}}^{\mathrm{RS}}(Q,\Gamma;\alpha,\kappa).
\label{eq:km-entropy-branch}
\end{equation}
The zero of this RS branch gives the Krauth--M\'ezard capacity prediction,
\begin{equation}
\alpha_{\mathrm{KM}}(\kappa)
=
\sup\{\alpha:\Sigma_{\mathrm{KM}}(\alpha,\kappa)>0\}.
\label{eq:km-capacity-def-revised}
\end{equation}

The hard formulas are obtained by this specialization of the scalar constraint recursion. Their finite-step extensions are calculated in the next section.

\section{One-step and finite-step formulas}
\label{sec:onestep}

We now write the finite-step CCM functional in the form used for one-step and finite-step computations.
The finite-step functional is
\begin{equation}
\mcG_{\mathrm{perc}}^{\mathrm{CCM}}(\zeta,Q,\Gamma)
=
X_0(0)+\alpha Y_0(0)
-
\frac12
\sum_{\ell=0}^{R}
\zeta_\ell
\left(
Q_{\ell+1}\Gamma_{\ell+1}-Q_\ell\Gamma_\ell
\right).
\label{eq:onestep-start-functional}
\end{equation}
where the profiles \(Q\) and \(\Gamma\) are the constructive variance profiles of the two sides of the auxiliary bipartite model.  The notation \(\Gamma\) is kept throughout the derivation.  Only after CCM calculation is complete do we compare it with the conventional perceptron notation, where the same scalar is usually denoted by \(\widehat q\).

\subsection{One-step profile}

The one-step profile has two nontrivial constructive levels.  On the spin side we use
\begin{equation}
Q_0=0,
\qquad
Q_1=Q_{\mathrm{out}},
\qquad
Q_2=Q_{\mathrm{in}},
\qquad
Q_3=1,
\end{equation}
\begin{equation}
0\le Q_{\mathrm{out}}\le Q_{\mathrm{in}}<1.
\end{equation}
On the auxiliary side we use
\begin{equation}
\Gamma_0=0,
\qquad
\Gamma_1=\Gamma_{\mathrm{out}},
\qquad
\Gamma_2=\Gamma_{\mathrm{in}},
\qquad
\Gamma_3=\Gamma_{\mathrm{in}},
\end{equation}
\begin{equation}
0\le \Gamma_{\mathrm{out}}\le \Gamma_{\mathrm{in}}.
\end{equation}
The Parisi averaging parameters are
\begin{equation}
\zeta_0=0,
\qquad
\zeta_1=m,
\qquad
\zeta_2=1,
\qquad
0<m\le1.
\end{equation}
The words ``out'' and ``in'' are labels for the constructive variance levels.  They are not additional assumptions on the microscopic configuration space.

\subsection{Site term}

The site terminal is defined as follows:
\begin{equation}
X_3(u)=\log2\cosh u.
\end{equation}
Since \(\Gamma_3-\Gamma_2=0\), the top step carries no site variance and
\begin{equation}
X_2(u)=X_3(u)=\log2\cosh u.
\end{equation}
The one-step layer has variance \(\Gamma_{\mathrm{in}}-\Gamma_{\mathrm{out}}\) and parameter \(m\). Hence
\begin{equation}
X_1(u)
=
\frac1m
\log
\E_{Z_1}
\left[
2\cosh
\left(
 u+
\sqrt{\Gamma_{\mathrm{in}}-\Gamma_{\mathrm{out}}}\,Z_1
\right)
\right]^m.
\end{equation}
The bottom step has \(\zeta_0=0\), so it is an ordinary Gaussian expectation.  Thus
\begin{equation}
I_{\mathrm{site}}
:=X_0(0)
=
\E_{Z_0}
\frac1m
\log
\E_{Z_1}
\left[
2\cosh
\left(
\sqrt{\Gamma_{\mathrm{out}}}\,Z_0+
\sqrt{\Gamma_{\mathrm{in}}-\Gamma_{\mathrm{out}}}\,Z_1
\right)
\right]^m.
\label{eq:onestep-site-term}
\end{equation}
This is the one-step Ising site contribution generated by the auxiliary-side variance profile.

\subsection{Soft constraint term}

For the soft channel the terminal is as follows:
\begin{equation}
Y_3(u)=\log U(u).
\end{equation}
The top step has variance \(1-Q_{\mathrm{in}}\) and parameter \(1\). Therefore
\begin{equation}
Y_2(u)
=
\log
\E_{Z_2}
U\left(
 u+
\sqrt{1-Q_{\mathrm{in}}}\,Z_2
\right).
\end{equation}
The one-step layer has variance \(Q_{\mathrm{in}}-Q_{\mathrm{out}}\) and parameter \(m\):
\begin{equation}
Y_1(u)
=
\frac1m
\log
\E_{Z_1}
\left[
\E_{Z_2}
U\left(
 u+
\sqrt{Q_{\mathrm{in}}-Q_{\mathrm{out}}}\,Z_1+
\sqrt{1-Q_{\mathrm{in}}}\,Z_2
\right)
\right]^m.
\end{equation}
The bottom step averages over the common Gaussian component of variance \(Q_{\mathrm{out}}\). Hence
\begin{equation}
I_{\mathrm{constr}}^{\mathrm{soft}}
:=Y_0(0)
=
\E_{Z_0}
\frac1m
\log
\E_{Z_1}
\left[
\E_{Z_2}
U\left(
\sqrt{Q_{\mathrm{out}}}\,Z_0+
\sqrt{Q_{\mathrm{in}}-Q_{\mathrm{out}}}\,Z_1+
\sqrt{1-Q_{\mathrm{in}}}\,Z_2
\right)
\right]^m.
\label{eq:onestep-soft-constraint-term}
\end{equation}
This is the one-step constraint contribution.  The innermost expectation is the residual Gaussian smoothing of the perceptron channel at variance \(1-Q_{\mathrm{in}}\).

\subsection{Correction term}

The correction in \eqref{eq:onestep-start-functional} is as follows:
\begin{equation}
-
\frac12
\sum_{\ell=0}^{2}
\zeta_\ell
\left(
Q_{\ell+1}\Gamma_{\ell+1}-Q_\ell\Gamma_\ell
\right).
\end{equation}
The first term vanishes because \(\zeta_0=0\).  The second term is
\begin{equation}
-
\frac12
m
\left(
Q_{\mathrm{in}}\Gamma_{\mathrm{in}}
-
Q_{\mathrm{out}}\Gamma_{\mathrm{out}}
\right),
\end{equation}
and the final term is therefore
\begin{equation}
-
\frac12
\left(
\Gamma_{\mathrm{in}}-Q_{\mathrm{in}}\Gamma_{\mathrm{in}}
\right)
=
-
\frac12
(1-Q_{\mathrm{in}})\Gamma_{\mathrm{in}}.
\end{equation}
Thus the one-step correction is
\begin{equation}
-
\frac12
\left[
 m
\left(
Q_{\mathrm{in}}\Gamma_{\mathrm{in}}
-
Q_{\mathrm{out}}\Gamma_{\mathrm{out}}
\right)
+
(1-Q_{\mathrm{in}})\Gamma_{\mathrm{in}}
\right].
\label{eq:onestep-correction-term}
\end{equation}
This term is the bipartite analogue of the usual Parisi correction.  The quadratic SK scale is replaced by the product scale \(Q\Gamma\).

\subsection{One-step soft functional}

Combining \eqref{eq:onestep-site-term}, \eqref{eq:onestep-soft-constraint-term}, and \eqref{eq:onestep-correction-term}, the one-step soft perceptron functional is
\begin{equation}
\mcG_{\mathrm{soft}}^{1\mathrm{RSB}}
=
I_{\mathrm{site}}
+
\alpha I_{\mathrm{constr}}^{\mathrm{soft}}
-
\frac12
\left[
 m
\left(
Q_{\mathrm{in}}\Gamma_{\mathrm{in}}
-
Q_{\mathrm{out}}\Gamma_{\mathrm{out}}
\right)
+
(1-Q_{\mathrm{in}})\Gamma_{\mathrm{in}}
\right].
\label{eq:onestep-soft-functional-expanded}
\end{equation}
The CCM computation supplies the finite-step functional, and stationary equations characterize its interior candidates. The soft variational prescription is recalled in Section~\ref{sec:openai-comparison}; boundary candidates are tested by admissible one-sided variations.

\subsection{Hard one-step functional}

For the hard channel, the \(U\) function is specialized to
\begin{equation}
U_\kappa(u)=\one\{u\ge\kappa\},
\end{equation}
the residual smoothing gives
\begin{equation}
\E_{Z_2}
U_\kappa
\left(
 u+
\sqrt{1-Q_{\mathrm{in}}}\,Z_2
\right)
=
H\left(
\frac{\kappa-u}{\sqrt{1-Q_{\mathrm{in}}}}
\right).
\end{equation}
Therefore we find
\begin{equation}
I_{\mathrm{constr}}^{\mathrm{hard}}
=
\E_{Z_0}
\frac1m
\log
\E_{Z_1}
\left[
H\left(
\frac{
\kappa
-
\sqrt{Q_{\mathrm{out}}}\,Z_0
-
\sqrt{Q_{\mathrm{in}}-Q_{\mathrm{out}}}\,Z_1
}{
\sqrt{1-Q_{\mathrm{in}}}
}
\right)
\right]^m.
\label{eq:onestep-hard-constraint-term}
\end{equation}
The hard one-step functional is
\begin{equation}
\mcG_{\mathrm{hard}}^{1\mathrm{RSB}}
=
I_{\mathrm{site}}
+
\alpha I_{\mathrm{constr}}^{\mathrm{hard}}
-
\frac12
\left[
 m
\left(
Q_{\mathrm{in}}\Gamma_{\mathrm{in}}
-
Q_{\mathrm{out}}\Gamma_{\mathrm{out}}
\right)
+
(1-Q_{\mathrm{in}})\Gamma_{\mathrm{in}}
\right].
\label{eq:onestep-hard-functional-expanded}
\end{equation}
With the a posteriori identification
\begin{equation}
Q_{\mathrm{out}}=q_0,
\qquad
Q_{\mathrm{in}}=q_1,
\qquad
\Gamma_{\mathrm{out}}=\widehat q_0,
\qquad
\Gamma_{\mathrm{in}}=\widehat q_1,
\end{equation}
this is the standard one-step Gardner--Krauth--M\'ezard form.

\subsection{Collapse to the replica-symmetric formula}

The one-step formula reduces to the replica-symmetric formula when the intermediate variance increments vanish: we take
\begin{equation}
Q_{\mathrm{out}}=Q_{\mathrm{in}}=Q,
\qquad
\Gamma_{\mathrm{out}}=\Gamma_{\mathrm{in}}=\Gamma.
\end{equation}
Then \eqref{eq:onestep-site-term} becomes as follows:
\begin{equation}
I_{\mathrm{site}}=\E_Z\log2\cosh(\sqrt\Gamma Z),
\end{equation}
and \eqref{eq:onestep-soft-constraint-term} becomes as follows:
\begin{equation}
I_{\mathrm{constr}}^{\mathrm{soft}}
=
\E_Z\log\E_{Z'}U(\sqrt QZ+\sqrt{1-Q}Z').
\end{equation}
Finally, the correction term becomes
\begin{equation}
-
\frac12(1-Q)\Gamma.
\end{equation}
Thus the one-step soft expression collapses to the RS Gardner expression.  In the hard channel, the same collapse gives the RS Krauth--M\'ezard expression.

\subsection{General finite-step form}

For general profiles
\begin{equation}
0=Q_0\le Q_1\le\cdots\le Q_{R+1}=1,
\qquad
0=\Gamma_0\le\Gamma_1\le\cdots\le\Gamma_R,
\qquad
\Gamma_{R+1}:=\Gamma_R,
\end{equation}
\begin{equation}
0=\zeta_0\le\zeta_1\le\cdots\le\zeta_R=1,
\end{equation}
the finite-step formula remains
\begin{equation}
\mcG_{\mathrm{soft}}^{(R)}
=
X_0(0)+\alpha Y_0(0)
-
\frac12
\sum_{\ell=0}^{R}
\zeta_\ell
\left(
Q_{\ell+1}\Gamma_{\ell+1}-Q_\ell\Gamma_\ell
\right).
\label{eq:finite-step-soft-expanded}
\end{equation}
The site recursion is as follows:
\begin{equation}
X_{R+1}(u)=\log2\cosh u,
\qquad
X_\ell=T_{\zeta_\ell,\Gamma_{\ell+1}-\Gamma_\ell}X_{\ell+1}.
\end{equation}
The soft constraint recursion is
\begin{equation}
Y_{R+1}(u)=\log U(u),
\qquad
Y_\ell=T_{\zeta_\ell,Q_{\ell+1}-Q_\ell}Y_{\ell+1}.
\end{equation}
For the hard channel with positive residual variance, the final smoothed terminal is
\begin{equation}
Y_R^{\mathrm{hard}}(u)
=
\log
H\left(
\frac{\kappa-u}{\sqrt{1-Q_R}}
\right),
\qquad Q_R<1.
\end{equation}
We then run the same recursion down to level zero. The finite-step hard functional is
\begin{equation}
\mcG_{\mathrm{hard}}^{(R)}
=
X_0(0)+\alpha Y_0^{\mathrm{hard}}(0)
-
\frac12
\sum_{\ell=0}^{R}
\zeta_\ell
\left(
Q_{\ell+1}\Gamma_{\ell+1}-Q_\ell\Gamma_\ell
\right).
\label{eq:finite-step-hard-expanded}
\end{equation}
The formulas above give the finite-step Gardner functional in CCM variables.

\section{Analytic checks}

We record the checks that fix the normalization of the finite-step functional and the coefficient of the bipartite correction term. The functional is
\begin{equation}
\mcG_{\mathrm{perc}}^{\mathrm{CCM}}
=
X_0(0)+\alpha Y_0(0)
-
\frac12
\sum_{\ell=0}^{R}
\zeta_\ell
\left(
Q_{\ell+1}\Gamma_{\ell+1}-Q_\ell\Gamma_\ell
\right).
\label{eq:checks-main-functional-expanded}
\end{equation}

\subsection{Zero constraint density and constant channels}

If \(M_N=0\), then \(Z_N=2^N\). The constraint term disappears. The auxiliary profile is selected at \(\Gamma_\ell=0\), the site recursion gives \(X_0(0)=\log2\), and the correction vanishes. Hence
\begin{equation}
\mcG_{\mathrm{perc}}^{\mathrm{CCM}}=\log2.
\end{equation}

If \(U\equiv c>0\), then \(Z_N=2^N c^{M_N}\), and
\begin{equation}
\frac1N\log Z_N
=\log2+\frac{M_N}{N}\log c
\longrightarrow\log2+\alpha\log c.
\end{equation}
The CCM constraint recursion has terminal \(\log c\), so \(Y_0(0)=\log c\), while the auxiliary profile and the correction vanish. The same limiting value is obtained.

\subsection{Annealed branch}

For a fixed spin configuration, the field
\begin{equation}
S_\mu(\sigma)=\frac1{\sqrt N}\sum_i\xi_i^\mu\sigma_i
\end{equation}
has standard Gaussian law. Therefore
\begin{equation}
\frac1N\log \E Z_N
=
\log2+\frac{M_N}{N}\log\E_ZU(Z)
\longrightarrow\log2+\alpha\log\E_ZU(Z).
\label{eq:checks-annealed-expanded}
\end{equation}
The CCM reproduces this branch by taking the trivial non-final spin profile and the zero auxiliary profile. Then \(X_0(0)=\log2\), the constraint recursion contains only the final residual Gaussian variance, \(Y_0(0)=\log\E_ZU(Z)\), and the correction is zero.

\subsection{Linear channel}

Take the exponential form of the \(U\) function:
\begin{equation}
U(u)=e^{hu}.
\end{equation}
The original model becomes an Ising model in an independent Gaussian random field:
\begin{equation}
\prod_{\mu=1}^{M_N}e^{hS_\mu(\sigma)}
=
\exp\left\{
\sum_{i=1}^{N}H_i\sigma_i
\right\},
\qquad
H_i=
\frac h{\sqrt N}\sum_{\mu=1}^{M_N}\xi_i^\mu.
\end{equation}
Thus \(H_i\) has variance \(\alpha h^2+o(1)\), and the exact limiting pressure is
\begin{equation}
\E_Z\log2\cosh(\sqrt{\alpha h^2}\,Z).
\label{eq:checks-linear-exact-expanded}
\end{equation}
The RS CCM functional gives
\begin{equation}
\mcG_{\mathrm{RS}}(Q,\Gamma)
=
\E_Z\log2\cosh(\sqrt\Gamma Z)
+
\alpha
\E_Z\log\E_{Z'}e^{h(\sqrt QZ+\sqrt{1-Q}Z')}
-
\frac12\Gamma(1-Q).
\end{equation}
The constraint term is
\begin{equation}
\E_Z\log\E_{Z'}e^{h(\sqrt QZ+\sqrt{1-Q}Z')}
=
\frac12h^2(1-Q).
\end{equation}
Hence, the formula becomes
\begin{equation}
\mcG_{\mathrm{RS}}(Q,\Gamma)
=
\E_Z\log2\cosh(\sqrt\Gamma Z)
+
\frac12(\alpha h^2-\Gamma)(1-Q).
\end{equation}
Stationarity in \(Q\) gives \(\Gamma=\alpha h^2\). Substitution gives exactly \eqref{eq:checks-linear-exact-expanded}. This check fixes the normalization of the correction term.

\section{Relation with the OpenAI variational formula}
\label{sec:openai-comparison}

The finite-step functional \eqref{eq:ppp-final-functional} agrees with the variational formulation of the Ising random perceptron given by OpenAI~\cite{OpenAIIsingPerceptron2026}. To compare the two expressions, identify the log-potential and the step paths of that reference through
\begin{equation}
\begin{gathered}
f=\log U,\qquad
q(u)=Q_{\ell+1},\qquad
h(u)=\Gamma_{\ell+1},\\
\zeta_\ell<u<\zeta_{\ell+1},
\qquad \ell=0,\ldots,R-1.
\end{gathered}
\label{eq:openai-path-dictionary}
\end{equation}
Here the two paths describe the spin and conjugate-field profiles. The Gaussian patterns have the same normalization; Ref.~\cite{OpenAIIsingPerceptron2026} uses the uniform probability measure on the Ising configurations, whereas our partition function uses counting measure. The corresponding pressures therefore differ by the logarithm of two.

Write \(V_f[q]\) and \(\ell[h]\) for the pattern and compensated one-site functionals of Ref.~\cite[Definition~2.1]{OpenAIIsingPerceptron2026}. Their Gaussian recursions have the same variance increments and averaging parameters as \eqref{eq:site-recursion} and \eqref{eq:constraint-recursion}. The pattern recursion includes the final residual variance, while our last auxiliary increment is zero. Comparing the terminal functions gives
\begin{equation}
V_f[q]=Y_0(0),
\qquad
\ell[h]=X_0(0)-\log2-\frac{\Gamma_R}{2}.
\label{eq:openai-recursion-dictionary}
\end{equation}
The second identity accounts for the deterministic subtraction in the field terminal of Ref.~\cite{OpenAIIsingPerceptron2026} and the factor two in our site terminal.

Using the endpoint conventions of our profiles, summation by parts rewrites the bipartite correction as
\begin{equation}
\sum_{\ell=0}^{R}\zeta_\ell\Delta(Q\Gamma)_\ell
=
\Gamma_R-\int_0^1q(u)h(u)\dd u.
\label{eq:openai-correction-dictionary}
\end{equation}
Substitution into \eqref{eq:ppp-final-functional} yields
\begin{equation}
\mcG_{\mathrm{perc}}^{\mathrm{CCM}}(\zeta,Q,\Gamma)-\log2
=
\alpha V_f[q]+\ell[h]
+\frac12\int_0^1q(u)h(u)\dd u.
\label{eq:openai-functional-identity}
\end{equation}
The right-hand side is the objective in Ref.~\cite[Definitions~2.1--2.2]{OpenAIIsingPerceptron2026} before taking the supremum over the conjugate field. Thus the functional forms coincide for every finite hierarchy, including the site, constraint and correction contributions.

Condition \eqref{eq:auxiliary-integrability} places our positive auxiliary channels in the integrable class of the Ref.~\cite[Theorem~8.3]{OpenAIIsingPerceptron2026}. For the positive channels covered by Theorems~2.3 and~8.3 of that reference, the thermodynamic prescription takes the supremum over the conjugate field at fixed spin path, followed by the infimum over the spin path. The two paths admit independent partitions and are evaluated on a common refinement. This identifies the CCM expression with the same variational objective throughout the channel classes covered by those results.

\section{Discussion}

The positive auxiliary representation turns the Ising perceptron into a dense bipartite Gaussian model with an Ising layer and a continuous auxiliary layer, which makes the blanket decomposition directly applicable: the blanket shells are differences of bipartite rectangles, the core comparison \eqref{eq:core-pressure-bound}, under \eqref{eq:auxiliary-integrability}, gives the interface-only pressure on refinement, and the interface produces the constructive site and constraint fields.

The scalar cavity functional has three contributions: the site contribution \(\log\langle2\cosh x\rangle\), the constraint contribution \(\alpha\log\langle U(s)\rangle\), and the correction \(-\log\langle e^y\rangle\). The constructive readings of the three marks are governed by \(\Gamma\), \(Q\), and \(Q\Gamma\), respectively. CCM identifies these variables with the Aizenman--Sims--Starr ROSt variables~\cite{SimplifiedParisiIII}, and Parisi's recursive averaging gives
\begin{equation}
\mcG_{\mathrm{perc}}^{\mathrm{CCM}}
=
X_0(0)+\alpha Y_0(0)
-
\frac12
\sum_{\ell=0}^{R}
\zeta_\ell
\left(
Q_{\ell+1}\Gamma_{\ell+1}-Q_\ell\Gamma_\ell
\right).
\end{equation}

The replica-symmetric specialization recovers Gardner's formula after the identification \(Q=q\), \(\Gamma=\widehat q\). Specializing the constraint terminal to the hard channel gives the Gaussian tail \(H\) and the Krauth--M\'ezard functional, with the RS capacity prediction obtained from the selected entropy branch. The very same recursion gives the one-step and finite-step forms.

The comparison in Section~\ref{sec:openai-comparison} shows that the CCM calculation yields the same finite-step functional as the variational formulation of OpenAI~\cite{OpenAIIsingPerceptron2026}. For the positive auxiliary channels satisfying \eqref{eq:auxiliary-integrability}, that work selects the thermodynamic value by the supremum over the conjugate field followed by the infimum over the spin profile. The present calculation obtains this functional through the constructive cavity variables and Parisi's recursive averaging. The same organization suggests extensions to generalized linear models, committee machines, and multilayer perceptrons, where the channel determines the constraint recursion and the interaction geometry determines the correction scale.

\section*{Acknowledgments}\label{sec:acknowledgments}
We thank Nicola Kistler (Goethe University Frankfurt) and ChatGPT (OpenAI) for important discussions and suggestions.

\section*{Declaration of Generative AI}
Research, writing, and proofreading of the present manuscript were performed with AI assistance (GPT-6 Astra) under strict human guidance and scrutiny. We independently checked all mathematical arguments and take full responsibility for the content of the publication.

\urlstyle{same}

\end{document}